\documentclass{aa}

\usepackage{natbib}
\usepackage{graphicx}

\usepackage{epsfig}
\usepackage{amssymb}
\usepackage{amsmath}
\usepackage{txfonts}
\bibpunct{(}{)}{,}{}{}{,}

\begin{document}
\title{Magnetic field strength distribution of magnetic bright points inferred from  filtergrams and spectro-polarimetric data}
\titlerunning{Magnetic field strength distribution of MBPs}

\author{D. Utz
\inst {1, 2}
\and J. Jur\v{c}\'{a}k
\inst {3}
\and A. Hanslmeier
\inst {2, 4}
\and R. Muller
\inst 4
\and A. Veronig
\inst 2
\and O. K{\"u}hner
\inst 2}

\institute{Instituto de Astrof\'{i}sica de Andaluc\'{i}a (CSIC), Apdo. de Correos 3004, 18080 Granada, Spain
\and IGAM/Institute of Physics, University of Graz, Universit{\"a}tsplatz 5, 8010 Graz, Austria
\and Astronomical Institute of the Academy of Sciences, Fricova 298, 25 165 Ond\v{r}ejov, Czech Republic
\and Laboratoire d\'{}Astrophysique de Toulouse et Tarbes, UMR5572, CNRS et Universit{\'e} Paul Sabatier Toulouse 3, 57 avenue d\'{}Azereix, 65000 Tarbes France}

\date{Received 15 March 2011 /
Accepted 19 March 2013 }

\abstract{Small scale magnetic fields can be observed on the Sun in G-band filtergrams as MBPs (magnetic bright points) or identified in spectro-polarimetric measurements due to enhanced signals of Stokes profiles. These magnetic fields and their dynamics play a crucial role in understanding the coronal heating problem and also in surface dynamo models. MBPs can theoretically be described to evolve out of a patch of a solar photospheric magnetic field with values below the equipartition field strength by the so-called convective collapse model. After the collapse, the magnetic field of MBPs reaches a higher stable magnetic field level.
} {The magnetic field strength distribution of small scale magnetic fields as seen by MBPs is inferred. Furthermore, we want to test the  model of convective collapse and the theoretically predicted stable value of about 1300 G.
} {We used four different data sets of high-resolution Hinode/SOT observations that were recorded simultaneously with the broadband filter device (G-band, Ca II-H) and the spectro-polarimeter. To derive the magnetic field strength distribution of these small scale features, the spectropolarimeter (SP) data sets were treated by the Merlin inversion code. The four data sets comprise different solar surface types: active regions (a sunspot group and a region with pores), as well as quiet Sun.
} {In all four cases the obtained magnetic field strength distribution of MBPs is similar and shows peaks around 1300 G. This agrees well with the theoretical prediction of the convective collapse model. The resulting magnetic field strength distribution can be fitted in each case by a model consisting of log-normal components. The important parameters, such as geometrical mean value and multiplicative standard deviation, are similar in all data sets, only the relative weighting of the components is different.} {}


\maketitle

\section{Introduction}

The dynamics of the Sun's atmosphere are dominated by magnetic fields. These fields span several orders of magnitude in both field strength and in size \citep[see e.g.][]{1987ARA&A..25...83Z}. From extended features like sunspot groups (kG fields) to small active patches and pores, down to the smallest detectable elements, isolated single flux tubes. With newly installed observing facilities \citep[Hinode, NST, Sunrise;][]{2007SoPh..243....3K,2010AN....331..636C,2011SoPh..268....1B} and highly sophisticated methods like inversions of spectro-polarimetric data that are used more and more \citep[e.g. the SIR code, the Merlin code;][]{1992ApJ...398..375R,2007MmSAI..78..148L}, the detection limit is constantly improving for weak and small magnetic fields in the solar atmosphere. Therefore, features smaller in size can be studied as can those weaker in magnetic field strength. This has led in recent years to a dramatic change in our understanding of the quiet Sun's magnetic fields. The magnetic fields of the quiet Sun (which exclude active regions) comprise network and intranetwork fields. The network fields spatially correspond to the supergranular boundaries. Inside of these supergranules, the fields are called intranetwork fields \citep[see e.g.][]{1987ARA&A..25...83Z}.

While it was previously thought that the Sun's magnetic field in quiet regions is built up by strong vertical magnetic flux tubes in the intergranular lanes (between granules) and field-free regions in between (granules), recent studies have revealed that the field-free thought domains (intranetwork regions) are in fact filled up with magnetic fields with strengths up to hundreds of Gauss \citep[see e.g.][]{2007ApJ...670L..61O,2008ApJ...672.1237L,2010ApJ...723L.149D}. These intranetwork fields contain a large number of horizontal fields and can be observed above granules. In many cases, horizontal intranetwork fields emerge co-spatially with granules and can be interpreted as low-lying loops \citep[see e.g.][]{2010A&A...511A..14G}. 
However, there are also recent studies suggesting other interpretations of the new observations than strong horizontal magnetic fields \citep[see e.g.:][]{2009ApJ...701.1032A,2010A&A...517A..37S,2011A&A...527A..29B}.
\begin{figure*}

\begin{center}
\includegraphics[width=0.8\textwidth]{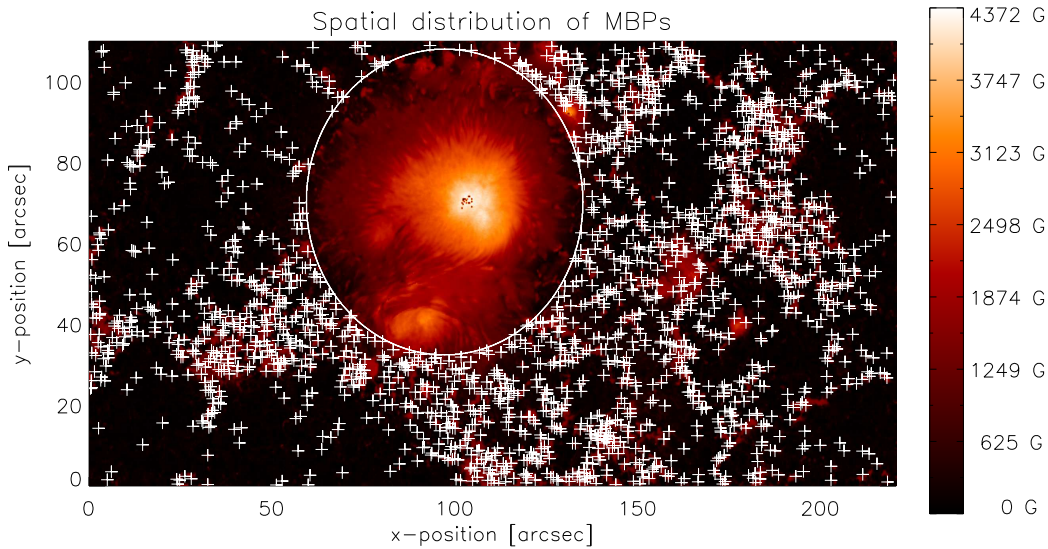}
\end{center}
\caption{Magnetic field strength map of the fully inverted SP data of data set I, together with the identified MBPs (indicated by crosses). The white circle outlines an exclusion zone around the sunspot; i.e. identified MBP features within this zone are not taken into account in the analysis. 
}
\label{figure1}
\end{figure*}

Such small-scale and weak magnetic fields can be observed in spectropolarimetric
data \citep[e.g.][]{2008A&A...490L..23B,2008ApJ...677L.145N,2009ApJ...700L.145V} or in magnetic-field sensitive filtergrams, such as the G-band \citep[e.g.][]{1998ApJ...506..439B,2004A&A...428..613B}. In such filtergrams, the magnetic field concentrations are identified as so-called magnetic bright points \citep[MBPs; see e.g.][]{1992Natur.359..307K,1993SoPh..144....1Y,2001ApJ...553..449B,2004ApJ...609L..91S,2009A&A...498..289U}. 
The G-band (centred on 430.5 nm) is primarily used for such investigations because it gives a better contrast between MBPs and the surrounding granulation than a continuum image would. The increased contrast is due to a partial evacuation of the flux tube. This leads to lower opacity, hence a deeper formation height in the atmosphere. As we can see deeper and therefore in a hotter atmospheric layer, the opacity in the G-band is further decreased owing to an increased dissociation rate of the contributing CH molecules. This gives rise to a higher contrast of MBPs in the G-band compared to the continuum.
 For more details about these processes, see \citet{2001A&A...372L..13S} and \citet{2003ApJ...597L.173S}. The relationship between the contrast and the magnetic field strength was confirmed by simulations like the one by \citet{2004A&A...427..335S}.

In the 1970s, \citet{1979SoPh...61..363S} developed a theoretical model for the formation process of MBPs, the so-called convective collapse model. The model suggests that when magnetic fields exceed a critical magnetic field strength (equipartition field strength\footnote{The field strength at which the magnetic pressure has reached a level sufficient to balance the kinetic pressure exerted on the mangetic field; so the magnetic field cannot get further compressed by the surrounding plasma flows.}), the plasma within the magnetic field cools down to form down drafts within the field, which finally leads to a collapse of the magnetic field, resulting in a small scale vertical flux tube. This flux tube is then visible as MBP.  The model suggests an equipartition magnetic field strength after the collapse of the magnetic field to a flux tube of roughly 1300 G.

In this paper, our aim is to study the magnetic field strength distribution of small scale magnetic fields observed as MBPs. A similar investigation was conducted by \citet{2008A&A...488.1101B}. These authors identified and analysed MBPs in G-band filtergram data to obtain the magnetic flux contained in these small scale kG field strength exhibiting features of the solar photosphere. For estimating of the magnetic field strength, they used a calibration curve of \citet{2004A&A...427..335S}. Nevertheless the direct estimation of the magnetic field strength distribution of MBPs on a larger scale and for different solar surface regions was not done before\footnote{There is a very interesting study by \citet{2007A&A...472..607B} of MBP properties in the sunspot moat. Among the studied parameters is the magnetic field strength and flux.} due to the lack of high-resolution co-temporal data sets comprising spectro-polarimetric data (suitable for inversions to gain magnetic field strength maps) and filtergram data (especially G-band data for the identification of MBPs). The Hinode satellite deployed in 2006 provides the right data for such an investigation since it combines a filtergram imager (G-band observations) with a high-resolution spectropolarimetric device (SP - spectro-polarimeter). Details about the Hinode mission and its SOT (Solar Optical Telescope) instrument can be found in \citet{2007SoPh..243....3K} and \citet{2008SoPh..249..167T}.


\begin{figure}

\begin{center}
\includegraphics[width=0.495\textwidth]{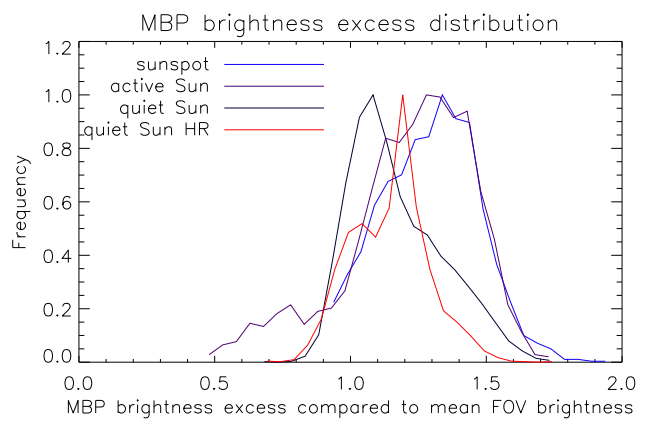}
\end{center}
\caption{The brightness excess distribution of MBPs identified in the four data sets. The brightness excess is defined as the brightness at the brightness barycentre position of an MBP divided by the mean brightness of the FOV (both taken from the G-band images).}
\label{figure2a}
\end{figure}

\section{Data sets}
We studied four different data sets comprising fully inverted SP spectropolarimeter data and G-band filtergrams. The SP inversions were carried out with the Merlin code\footnote{Detailed information about the Merlin code can be found in the internet under the following web URL: http://www.csac.hao.ucar.edu/csac/nextGeneration.jsp} (courtesey to Bruce Lites and co-workers). The data sets contain different typical solar features, suchas a sunspot group, a magnetically active region (pore), and quiet Sun. These data sets were chosen to a) cover different features of the Sun and b) feature nearly co-temporal filtergram and spectropolarimetric data (except data set I) with a high temporal cadence.
\begin{table*}
	\centering
	\caption{Fit parameters ($\sigma$ and $\mu$) and weighting coefficients ($w_1$ and $w_2$) derived for the magnetic field strength distribution of data set I (sunspot group; see Figs. \ref{figure1}, \ref{figure2}). The reduced $\chi^2$ value (describing the goodness of the fit) has a value of 3.9.}
		\begin{tabular}{ccccc}
		\hline
\hline
			description & weighting coefficient & relative weighting coefficient& mean value ($\mu$) & standard deviation ($\sigma$)\\
		 && [\%] & [G] & [G]\\
		 \hline
 log-normal distribution component I & $72000\pm2000$ & $66\pm2$ & $5.62\pm0.03$ &  $0.80\pm0.02$\\
		log-normal distribution component II& $37000\pm1500$ & $34\pm2$  &$7.13\pm0.01$ & $0.17\pm0.01$\\
		 	 \hline
		\end{tabular}
		\label{fit_1_values}
\end{table*}

\begin{figure}

\begin{center}
\includegraphics[width=0.48\textwidth]{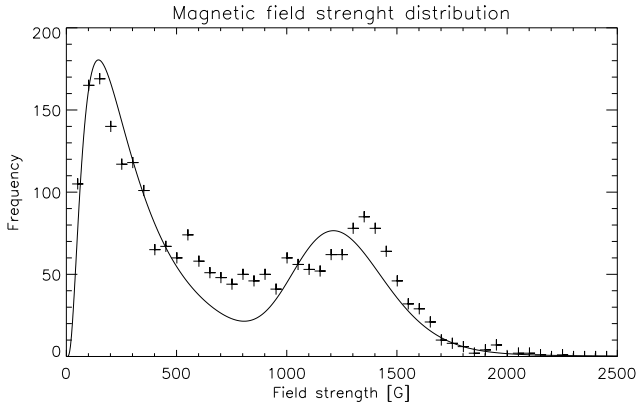}
\end{center}
\caption{The magnetic field strength distribution of MBP features gained by data set I (sunspot). The peak around 1300 G related to collapsed magnetic fields was fitted by a log-normal component yielding a geometric mean $\mu=(1250~\cdot|\div$~1.18) G. The reduced $\chi^2$ for the fit is 3.9. In total, 2392 MBP features were identified and used in the analysis which covered about 2 \% of the FOV.}
\label{figure2}
\end{figure}

\paragraph{Data Set I (sunspot group):} The data stem from 2006 December 11 and show a sunspot
group (see Fig. \ref{figure1})\footnote{Some pixels in the centre of the sunspot were not correctly inverted. This is most probably due to the strong magnetic fields in the centre of the sunspot. For our study this is of no concern as we are interested in the magnetic field strength of the MBP features which are located outside of the sunspot.}. A few days later this active region gained a reputation because it hosted the first X-class flare observed by Hinode \citep[see e.g.][]{2007PASJ...59S.779K,2008ApJ...672L..73J,2008ApJ...685..622A}. In
addition to the SP data, data of the
BFI (broad band filter imager) and NFI (narrow band filter
imager) device are available (G-band, Ca\,{\sc ii}\,H, Fe line). The main features of the SP data are as follows. The data were obtained by scanning the surface in the fast-map mode and cover the period from 17:00 UT to 18:03 UT. The totally scanned field of view (FOV) has a size of 295.2 arcsec by 162.3 arcsec. The centre of the FOV pointed to solar coordinates of $x=152.4$ arcsec, $y=-$96.1 arcsec. 
This position on the Sun corresponds to a heliocentric angle ($\theta$, inclination of vertical flux tubes to the line of sight) of 10.6$^{\circ}$ equaling a $\cos\theta$ value of 0.98. For the total FOV the variation in the inclination angles is within 4.1$^{\circ}$ to 17.4$^{\circ}$ ($\cos\theta$ value of 0.99 to 0.95), which is low enough that no special care has to be taken for projection effects. The inverted SP data (magnetogram) has a spatial sampling of 0.30 arcsec/pixel in the $x$-direction and of 0.32 arcsec/pixel in the $y$-direction. The corresponding filtergram data set had the following characteristics. The temporal resolution for the filtergram images is about seven minutes. The spatial sampling was reduced by onboard binning to 0.108 arcsec for the G-band and Ca\,{\sc ii}\,H filtergrams. In addition, Fe line filtergrams of the narrow band imager (NFI) instrument are available. These data have a spatial sampling of 0.16 arcsec/pixel. The filtergram time series covers the period from 15:32 UT to 18:58 UT with a different FOV of 221 arcsec by 110.5 arcsec. 

\paragraph{Data Set II (active region; pores):}
Data set II covers a small active region outside of the disc
centre (shifted to the north-east; $x=-330$ arcsec and $y=390$ arcsec). In this case the heliocentric angle has on average a value of 32.7$^{\circ}$ with a variation within 30.2$^{\circ}$ to 35.3$^{\circ}$ corresponding to a mean $\cos\theta$ of $0.84 \pm 0.2$. The exposures were taken over a time interval of about 18 minutes with a one-minute cadence for the filtergram exposures and a two-minutes cadence for the SP scans. The time series starts at 9:00 UT on 2009 June 2 and lasts until 9:20 UT. The data comprise Hinode
SOT/SP data fully inverted by the Merlin code (courtesy to B.
Lites et al.), G-band, Ca\,{\sc ii}\,H observations and Fe I line exposures of the NFI filter. The FOV of the filtergram exposures is about 80 by 90 $\mathrm{arcsec^2}$. The SP scans show an FOV of 9.45 arcsec by 81.15 arcsec and comprise ten fast map scans of the region. The spatial sampling is 0.295 arcsec/pixel. The Fe I line filtergrams are available in all four Stokes parameters ($I$,~$Q$,~$U$,~$V$) with a reduced FOV (compared to the BFI filtergrams) of 12.8 arcsec by 81.92 arcsec. The spatial sampling rate is higher than for data set I and amounts to 0.16 arcsec/pixel. The time series for the filtergram exposures covers a longer period, from 8:25 UT to 9:55 UT. In total, the filtergram data set comprise 90 exposures in each wavelength.

\paragraph{Data Set III (quiet Sun):}
Data set III was recorded on 2007 June 2. It covers a quiet Sun region near disc centre (about 200 arcsec
south of the disc centre). This gives a heliocentric angle for the FOV, hence also for the flux tubes in a range of 8.8$^{\circ}$ to 15.6$^{\circ}$ with a mean value of about 12.2$^{\circ}$. These values correspond to a mean $\cos\theta$ value of 0.98. For the total FOV the values range between 0.96 and 0.99. The data comprises SOT G-band,
Ca\,{\sc ii}\,H, and SP data. The SP data were inverted by the Merlin
code. The temporal coverage is about 2 minutes for the
magnetograms and about 30 seconds for the
filtergrams. In total this gives 26 fast-map SP scans of the region with a spatial sampling of 0.295 arcsec/pixel in the time span from 11:48 UT to 12:40 UT. The FOV of the SP data was 8.86 arcsec by 162.3 arcsec. Filtergram exposures were taken with a spatial sampling of 0.108 arcsec/pixel starting at the same time but lasting until 12:41. This gives 106 images for each of the two filters used. The FOV of the filtergrams is 55.8 arcsec by 111.6 arcsec.

\begin{figure}
\begin{center}
\includegraphics[width=0.378\textwidth]{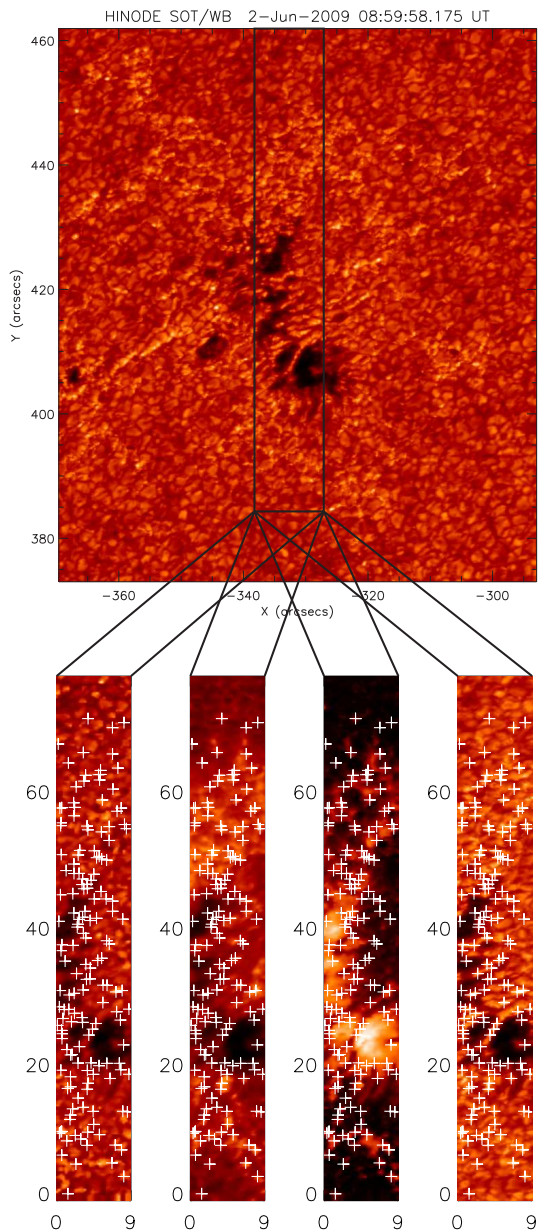}
\end{center}
\caption{Top: typical G-band filtergram of data set II (active region). The
region scanned by the SP instrument is marked by a rectangle. Bottom: available data and identified MBPs (marked by white crosses).
From left to right: whitelight image (obtained by inversion; SP data), Ca II-H filtergram
(chromosphere), magnetic field strength map (gained by inversion; SP data), and finally a G-band
filtergram. The range of the magnetic field strengths for this data is given as 14 to 2580 G.}
\label{figure3}
\end{figure}

\begin{figure}
\begin{center}
\includegraphics[width=0.48\textwidth]{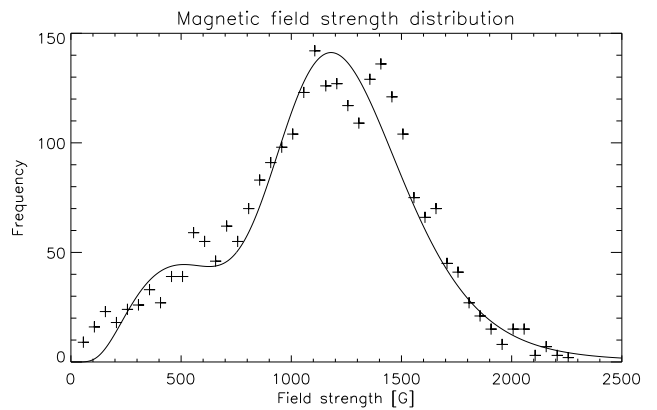}
\end{center}
\caption{The magnetic field strength distribution obtained for data set II (active Sun; pores). The fit was created by using two log-normal components. The first component assigned to MBPs yielded a geometric mean value of ($1260~\cdot|\div~1.25 $) G, while the second component is used for the background, which is rather weak. The reduced $\chi^2$ for the fit is 3.0. For this analysis, 2624 MBP features were taken into account and occupied a fraction of the FOV of about 6 \%.}
\label{figure4}
\end{figure}

\begin{table*}
	\centering
	\caption{Fit parameters for the MBP field strength distribution of data set II (active region; Fig. \ref{figure4}). The fit gives a reduced $\chi^2$ value of 3.0.}
		\begin{tabular}{ccccc}
		\hline
\hline
			description& weighting coefficient & relative weighting coefficient & mean value ($\mu$) & standard deviation ($\sigma$)\\
		 & &[\%] & [G] & [G]\\
		 \hline
upper panel; log-normal component I& $40000\pm6000$ & $31\pm5$  &$6.6\pm0.1$ & $0.59\pm0.05$\\
upper panel; log-normal component II & $87000\pm6000$ & $69\pm5$ & $7.14\pm0.01$ &  $0.22\pm0.01$\\
		 	 \hline
		\end{tabular}
		\label{table_active}
\end{table*}

\paragraph{Data Set IV (quiet Sun, HR data):}
The major difference of this data set to the previous ones is the higher spatial sampling. The instrument always obtains the same/or nearly the same diffraction limited data sets. Nevertheless, the resolution quality can be slightly changed by the spatial sampling rate \citep[see e.g.][]{2009Hvar}. While in the case of the other data sets, the data were binned two by two, data set IV comprises unbinned data. Therefore, the spatial sampling is doubled with 0.054 arcsec/pixel for filtergram images and 0.16 arcsec/pixel for the spectropolarimeter. This is the best spatial sampling Hinode can provide for the SOT instrument. The data were obtained on 2007 January 20 and were recorded between 6:54 UT and 7:58 UT, so they span roughly one hour of solar quiet Sun evolution. The G-band filtergram data were taken with a temporal resolution of 30 seconds, whereas the SP data have a cadence of roughly two minutes, so about four filtergram exposures belong to one SP slit scan. The FOV of the filtergram images is 400 pixels by 1024 pixels or 21.6 arcsec by 55.3 arcsec, respectively. The FOV of the SP data is 25 pixels by 512 pixels, corresponding to 3.7 arcsec by 81.9 arcsec. The SP instrument has been operated in normal mode compared to fast-map mode in the other data sets, which yields a higher spatial sampling of the data. The selected FOV (same FOV for BFI and SP) is about 3.7 arcsec by 55.1 arcsec. The SP device took in total 29 scans, whereas the filtergram imager took 128 exposures. The centre of the FOV had solar coordinates of $x=-$4.5 arcsec shifting during the time series to 5.6 arcsec and a $y=$2.9 arcsec. We can therefore assume, in very good approximation, that these data were all taken at the solar disc centre (heliocentric angle of 0$^{\circ}$ and $\cos\theta$ with a value of 1). The inclination of the flux tubes to the angle of sight therefore has (if at all) a negligible influence on the results.

\begin{table*}
	\centering
	\caption{Fit parameters ($\sigma$ and $\mu$) and weighting coefficients ($w_1$ and $w_2$) derived for the magnetic field strength distribution of data set III (quiet Sun; see Figs. \ref{figure5}, \ref{figure6}). The reduced $\chi^2$ goodness of fit parameter yielded a value of 8.3.}
		\begin{tabular}{ccccc}
		\hline
\hline
			description & weighting coefficient & relative weighting coefficient& mean value ($\mu$) & standard deviation ($\sigma$)\\
		 && [\%] & [G] & [G]\\
		 \hline	
	log normal distribution component I& $116000\pm2000$& $81\pm1$  &$5.69\pm0.01$ & $0.82\pm0.01$\\
		 log normal distribution component II & $28000\pm1000$ & $19\pm1$ & $7.208\pm0.003$ &  $0.085\pm0.002$\\
		 \hline
		\end{tabular}
		\label{fit_values_quiet_sun}
\end{table*}

\begin{figure}
\begin{center}
\includegraphics[width=0.4\textwidth]{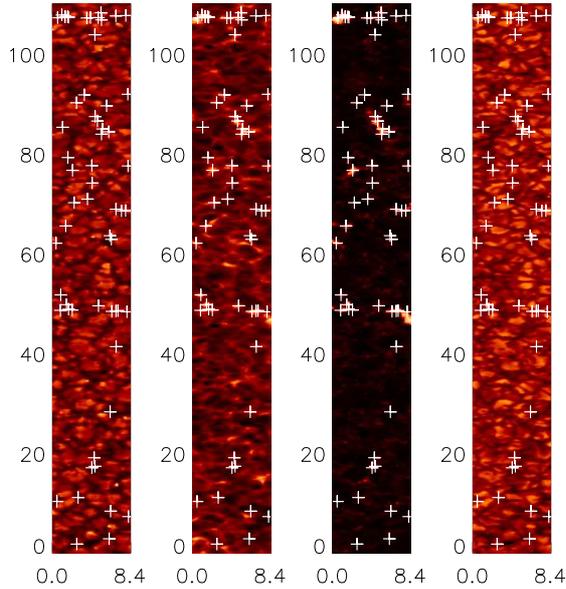}
\end{center}
\caption{From left to right: Continuum image stemming from SP data, CaII-H
(chromosphere), magnetic field strength map (inversion of SP data done by MERLIN code), G-band (photosphere). The crosses indicate
identified MBPs. It can be clearly seen that no significant magnetic field activity is
within the FOV (pores, sunspots). The magnetic field strengths in the data set vary between 3 and 1510 G.}
\label{figure5}
\end{figure}

\section{Analysis}
To analyse the magnetic field strength distribution of MBPs it is necessary to identify the MBPs correctly and then to determine their corresponding magnetic field strength. 
After obtaining beneficial data\footnote{These data sets must, on the one hand, comprise at least longitudinal magnetic field strength maps (coming e.g. from inversions; in the following, we refer to the magnetic field strength map somewhat loosely as a magnetogram) and on the other hand filtergram data of at least the G-band filter. Furthermore the temporal and spatial resolution should be high, and the magnetogram data to the filtergram data has to be as co-temporal as possible.} (see Sect. 2) from the Hinode data base, the next step is a careful alignment of the filtergram and magnetogram data sets. The G-band filtergrams have a different spatial and temporal sampling than the fully inverted level-2 spectro-polarimetric data. We overcame the different spatial sampling by stretching the magnetogram data sets to the corresponding spatial sampling of the filtergrams. This can be done in a first step by roughly expanding the magnetogram data set by the corresponding spatial sampling ratio. Since MBPs are very small-scale features, a rough alignment of $\pm 1$ arcsec would be insufficient. For a better alignment one can use the continuum intensity within the level-2 SP data sets. These continuum images show a granulation pattern similar to the G-band. By calculating cross-correlation coefficients, the expanded image can be aligned to the magnetogram on pixel and even subpixel scales. Finally, the data are cut to the same FOV. For details about high precision alignment and a corresponding algorithm for Hinode/SOT data, we refer to \citet{2010CEAB...34...31K}. The temporal alignment was done by taking the temporally closest filtergram to an SP scan. This gives temporal alignments with differences smaller than 0.5 minutes for data sets II to IV. In the case of data set I, we made just a spatial alignment of one filtergram to the complete SP scan. For problems evolving out of this we refer to the discussion section.

\begin{figure}
\begin{center}
\includegraphics[width=0.495\textwidth]{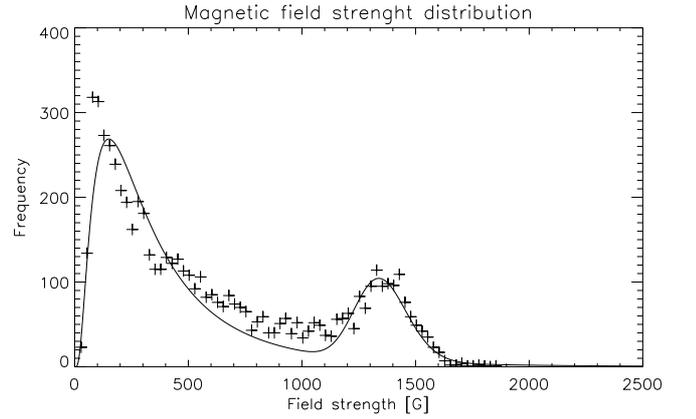}
\end{center}
\caption{Magnetic field strength distribution of the identified MBP features in a quiet solar FOV.
The hump related to MBPs is seen at (1350~$\cdot|\div$~1.09) G. The quality of the fit (solid line) is 8.3 (reduced $\chi^2$). The distribution was created by analysing 6253 identified MBP features covering about 1~\% of the FOV. }
\label{figure6}
\end{figure}

\begin{table*}
	\centering
	\caption{Fit parameters ($\sigma$ and $\mu$) and weighting coefficients ($w_1$ and $w_2$) derived for the magnetic field strength distribution of data set IV (quiet Sun, high-resolution data set; see Fig. \ref{figure7}). The measured goodness of fit in this case was 4.4.}
		\begin{tabular}{ccccc}
		\hline
\hline
			description & weighting coefficient & relative weighting coefficient& mean value ($\mu$) & standard deviation ($\sigma$)\\
		 && [\%] & [G] & [G]\\
		 \hline	
 log-normal distribution component I & $78000\pm1000$ & $93\pm2$ & $5.98\pm0.01$ &  $0.52\pm0.01$\\
		log-normal distribution component II& $6300\pm500$& $7\pm1$  & $7.05\pm0.01$ & $0.10\pm0.01$\\
		 \hline
		\end{tabular}
		\label{fit_values_quiet_sun_HR}
\end{table*}

After careful alignment the next step is to identify the interesting MBP features within the G-band data. This was done by the algorithm of \citet{2009A&A...498..289U}. The algorithm takes in a first step the G-band images and segments them into individual features. The idea of the segmentation is to follow the brightness contour of the features within an image from the brightest pixels down to the faintest. After applying of the segmentation the interesting MBP features are identified. The version applied here differs to the version of the cited paper (where the identification was based on the size of the features) in the sense that for the identification the local brightness gradient is considered. This means that for each segment the brightness ratio between the brightest pixel of a segment and an average brightness in a prescribed vicinity is calculated. If this ratio is high enough, the feature is identified as an MBP. If the ratio is close to 1, the segment belongs to a granule. The global brightness excess\footnote{Global means in this case the excess between an MBP feature to the mean G-band brightness of the total FOV. For the identification only local brightness excesses were considered.} distribution of MBPs is shown in Fig. \ref{figure2a}. The rest of the analysis is quite straightforward. After identifying the correct features, the barycentre position of the MBPs is calculated. The coordinates of the barycentre are then used to extract the corresponding magnetic field strength in the aligned magnetograms. The obtained magnetic field strengths are then represented in a histogram\footnote{The binsizes of the histograms were calculated by dividing the range of magnetic field strengths by an adequate number of representing classes. The number of classes can be estimated by calculating the square root of the number of data points, i.e. that the number of bins (and also the width of the bins) can vary from data set to data set depending on the total number of identified features. For the problem of correctly binning histograms see the treatise in \citet{salgado} comprising examples and clarifying the impact of changing binwidths.}, and the resulting distribution was finally fitted by a two-component log-normal distribution. The fit equation therefore is:
\begin{equation}
\begin{split}
&f(\sigma,\mu,x)=w_1 g(\sigma_{1},\mu_{1},x)+w_2 g(\sigma_{2},\mu_{2},x),
\end{split}
\end{equation}
where $g(\sigma,\mu,x)$ stands for the so-called log-normal distribution, $w_1$ and $w_2$ specify weighting coefficients. These weighting coefficients give the relative contribution of each component to the total function. The log-normal distribution $g(\sigma,\mu,x)$ is defined as
\begin{equation}
g(\sigma,\mu,x)=\frac{1}{x \sigma \sqrt{2 \pi}} \exp\left({-\frac{1}{2}\left(\frac{\log{x}-\mu}{\sigma}\right)^2}\right).
\end{equation}

The parameters are $\sigma$, a multiplicative standard deviation, and $\mu$ the geometric mean value. These parameters are somewhat different from the generally known $\mu$ and $\sigma$ parameters used in normal distributions as we employ log-normal distributions \citep[more information can be found in ][]{Eckhard}. 

The extraction of magnetic field information from magnetograms could lead to some errors for strongly inclined flux tubes. In our case the approach is appropriate because a) the data sets are close to the solar disc centre, hence producing low inclination angles, and b) MBPs are in general connected with strong vertical flux tubes.

\begin{table*}
	\centering
	\caption{Parameters for the log-normal component of the MBP magnetic field strength fit for all four data sets of our study. The mean value in a log-normal distribution is the geometrical mean and that the standard deviation is not addivitive but multiplicative.}
		\begin{tabular}{ccccc}
		\hline
\hline
			description & mean value fit ($\mu$) & mean value phys. & standard deviation fit ($\sigma$) & standard deviation as\\
		 & & [G] & & multiplicative factor\\
		 \hline
		 data set I (large FOV with sunspot) & $7.13\pm0.01$ & $1250\pm10$ & $0.17\pm0.01$ &  $1.18\pm0.02$\\
		 data set II (active Sun with pores) & $7.14\pm0.01$ & $1260\pm10$ & $0.22\pm0.01$ &  $1.25\pm0.01$\\
		 	 data set III (quiet Sun)  & $7.208\pm0.003$ & $1350\pm10$& $0.085\pm0.002$ & $1.08\pm0.01$\\
data set IV (quiet Sun; HR data) & $7.05\pm0.01$ & $1150\pm10$& $0.10\pm0.01$ & $1.10\pm0.01$\\
		 	 \hline
		\end{tabular}
		\label{mag_tab}
\end{table*}

\begin{figure}

\begin{center}
\includegraphics[width=0.48\textwidth]{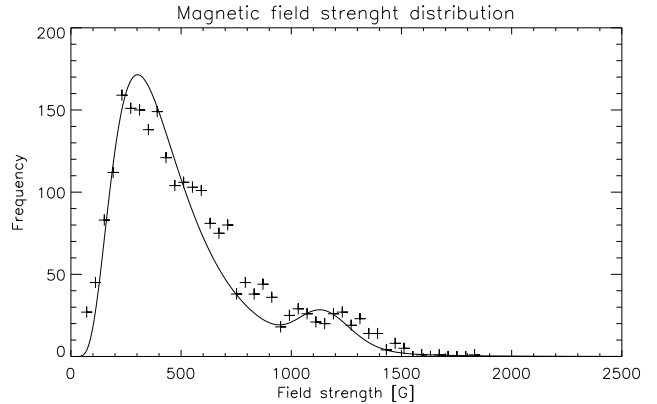}
\end{center}
\caption{Magnetic field strength distribution of the identified MBP features in the quiet solar FOV of the high spatial resolution data set (data set IV). The reduced $\chi^2$ for the fit is 4.4. The component related to MBPs is fitted at (1110~$\cdot|\div$~1.14 ) G. For this data set, 2614 identified MBP features were analysed, which covered about 1.4 \% of the total FOV.}
\label{figure7}
\end{figure}

\section{Results}
\paragraph{Data Set I (sunspot group):}The obtained magnetic field strength distribution for MBP features is shown in Fig. \ref{figure2}. In the figure caption of this figure and the following figures, the symbol $\cdot|\div$ means that the one $\sigma$ boundary can be gained by multiplying (higher boundary) and dividing (lower boundary) the geometric mean value by the multiplicative deviation. We use log-normal distributions and not the commonly used normal distributions for which the boundaries can be calculated by adding and subtracting the standard deviation. The fit values we derived are listed in Table \ref{fit_1_values}. The two fitted components correspond to one log-normal component that fits the magnetic field strength background signal and a second one that fits flux tubes with higher magnetic field strengths. This might be MBPs in their collapsed state; post-collapse MBPs. The background distribution can be explained by considering of MBPs before they undergo the collapse, during the collapse, and probably during the dissolution. Also a fraction of wrongly identified, non-MBP features might be taken into account. Finally the shape of this component is related to the dynamo process itself, creating magnetic fields up to the equipartition field strength, and to the way instrumental noise contributes via the inversion tools to the magnetograms. The more interesting component belongs to the collapsed flux tubes and is centred at 1250 G with a multiplicative standard deviation of 1.18 and a relative contribution of 34\%. This component can be attributed to the field strength of MBPs after the convective collapse. 1350 G is predicted by the convective collapse model e.g. in \citet{1979SoPh...61..363S}.

In the case of this data set one should have in mind that the SP instrument scans over the FOV. Such a scan needs about one hour for the actual size of the FOV. Therefore the left-hand side of the image shows the Sun's magnetic field at about 17:00 UT, whereas the right-hand side of the image shows the Sun's magnetic field at about 18:00 UT. We keep in mind that the evolution of small-scale fields happens in the range of minutes \citep[see e.g.:][]{2005A&A...441.1183D,2010A&A...511A..39U}. Therefore it is clear that the correlation between a single G-band filtergram and the magnetogram can only be statistical, i.e. that only a fraction of MBPs will be measured in the collapsed state with the correct magnetic field (giving rise to a relative contribution of 34\%), while for the others the magnetic field before or after their appearance will be measured (giving rise to a broad background log-normal distribution component of 66\%). For more details about the implications we refer to the discussion section.

\begin{figure}
\begin{center}
\includegraphics[width=0.48\textwidth]{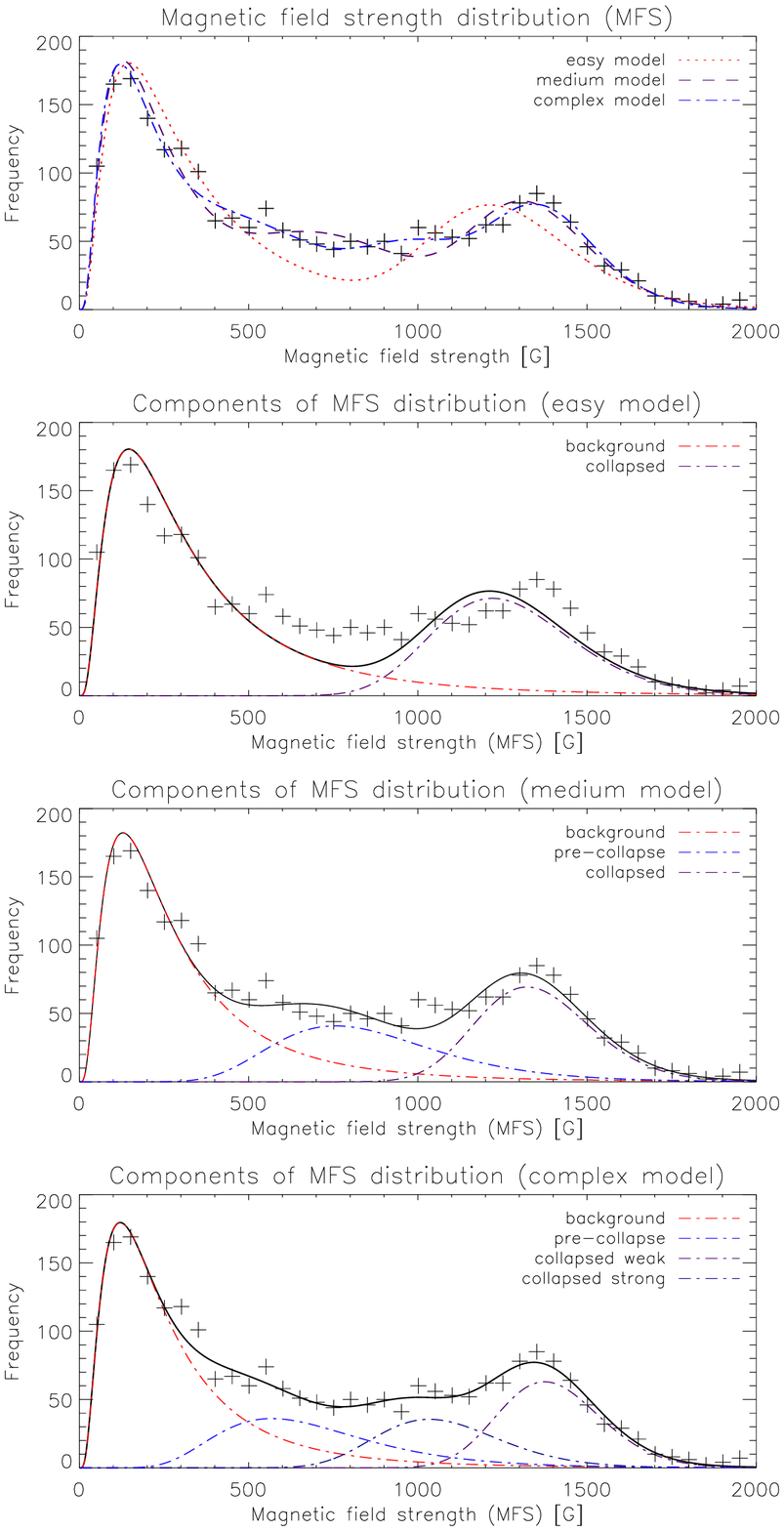}
\end{center}
\caption{Top panel: simplistic fit model (two log-normal components) applied to data set I (red, short dashed line) compared to a medium complex model consisting of 3 log-normal components (purple, long dashed line) and a sophisticated (4 log-normal components) model blue, short/long dashed line). Second panel: illustration of the components of the simplistic model. Third panel: components of the medium complex model. Fourth panel: components of the sophisticated fitting model, which assumes two strong magnetic field components around 1000 G and 1300 G, respectively. The reduced $\chi^2$ value for the simplistic model is 3.9 compared to 1.3 for the medium complex model and 0.8 for the most sophisticated one.}
\label{figure8}
\end{figure}

\begin{figure}
\begin{center}
\includegraphics[width=0.48\textwidth]{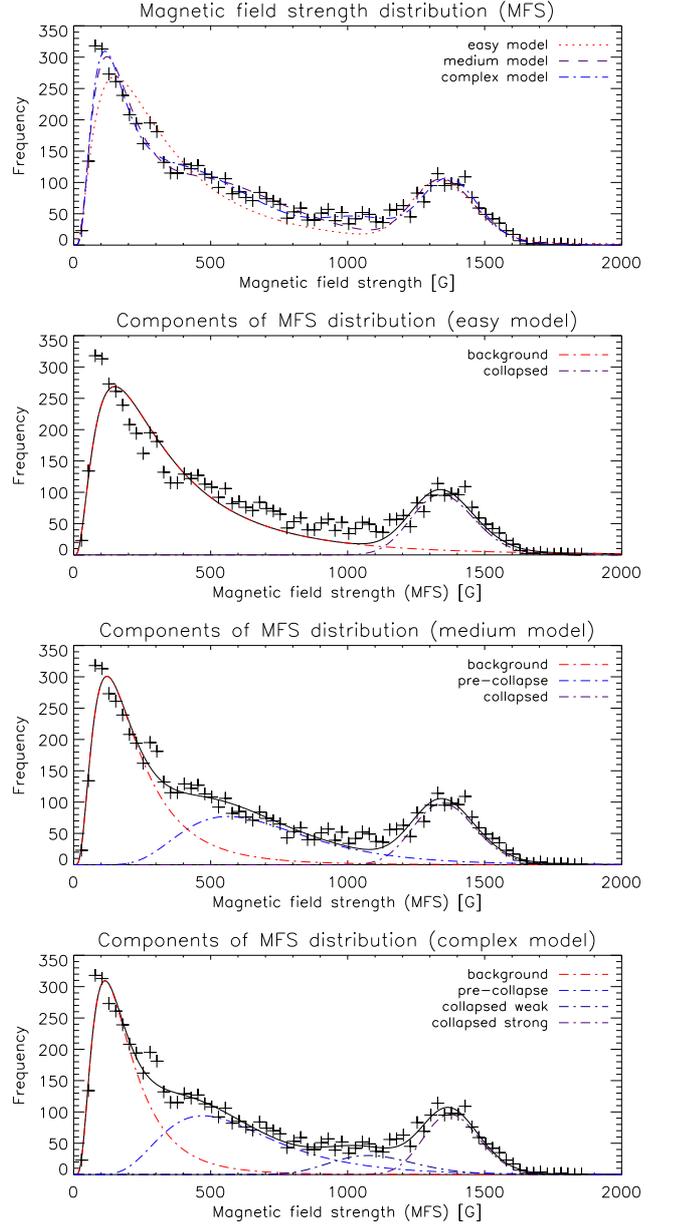}
\end{center}
\caption{Same as in Fig. \ref{figure8} but for data set III. It can be seen that the components are in position (but not strength; weighting) similar to the ones in Fig. \ref{figure8}. The goodness-of-fit values (reduced $\chi^2$) are 8.3 for the simplistic model (2 components; second panel) compared to 3.2 for the medium complex model (3 components; third panel) and 1.8 for the more sophisticated one (4 components; final panel).}

\label{figure9}
\end{figure}

\paragraph{Data Set II (active region; pores):} The main difficulty arises in the temporal and spatial alignment. The temporally closest subfields of the filtergrams corresponding to the scanned
FOV of the SP instrument were selected and aligned
(accuracy better than one pixel and two minutes, respectively). For an illustrative example see Fig. \ref{figure3}.
Figure \ref{figure4} shows the obtained magnetic field strength distribution
for this data set. The fit follows again the two-component log normal model. The geometrical mean for the MBP related component has a value of $(1260 \pm 10)$ G and a multiplicative standard deviation of 1.25. All obtained fit values are given in Table \ref{table_active}. Both components have similar slopes and similar mean values compared to the case of data set I (sunspot) and yet the relative contributions of background and collapsed flux tubes have significantly changed from 66\% to 31\% and 34\% to 69\%.

\paragraph{Data Set III (quiet Sun):}
Figure \ref{figure5} displays the different filtergram and magnetogram data of the used data set, and Fig. \ref{figure6} gives the measured magnetic field strength distribution for this case. The fit shown in Fig. \ref{figure6} was created by the same fit function as for data set I, i.e. two log-normal components. The main finding compared to the other data sets is that the relative importance of the convective collapse magnetic field strength peak about 1350 G is relatively weaker than in the other data sets. For data set III the relative weighting value derived is only 19\% compared to 34\% for data set I and 69\% for data set II. The peak is centred at 1350 G, which is a slightly higher value compared to the one obtained for data set I. The variance of the log-normal component is 1.09 (multiplicative variance of 9\%), which is by a factor of 2 smaller compared to data set I for which we derived a value of 1.18 (multiplicative variance of 18\%). For more details about the obtained fit values we refer to Table \ref{fit_values_quiet_sun}.

\paragraph{Data Set IV (quiet Sun; high-resolution data):}
The analysis of this data set differs from the analysis of the other data sets by the exclusion of MBP features with sizes smaller than at least four pixels. This is justified for two reasons: a) the smallest sizes are more likely to contain noise and/or are produced by noise (esepecially due to the higher spatial sampling) and b) a four-pixel size represents the same minimum size detectable in the other data sets (binning 2 by 2), hence an exclusion of smaller sizes improves the comparability between the different data sets. Figure \ref{figure7} shows the obtained magnetic field strength distribution for this data set and the results of the applied fitting model. As in the other data sets a significant log-normal component for collapsed flux tubes (MBPs) shows up at a value of (1150 $\pm$ 10) G for the high-resolution data. The multiplicative variance factor for this component is 1.10. The background log-normal component shows values of $\mu=5.98$ and $\sigma=0.52$. For more details about the obtained parameters we refer to Table \ref{fit_values_quiet_sun_HR}.

\paragraph{Summary:} 
In all four data sets we were able to fit the obtained magnetic field strength distribution of MBPs with a model consisting of two log-normal distributions. One component was needed for the background (non-collapsed flux tubes) and the second component for the collapsed state of flux tubes. This
type of distribution is known to be formed in agglomeration
processes like the growing of molecules on substrates, e.g. size distributions \citep[see][]{Eckhard}. In general, a random variable or measured quality shows this kind of distribution if the underlying process, which leads to the formation of the value, is created by multiplicative processes rather than by additive processes. Therefore the obtained fit
parameters and shapes of the distributions could yield a statistically
significant insight into the magnetic field generation (dynamo) process.

For detailed values of the fits see Table \ref{mag_tab}. The values reveal that the spread of the geometric mean for the MBPs ranges from 1110~G to 1350~G (data set IV and data set III, respectively). This can be solved by the following idea, on which we elaborate more in the next section, that we will indeed not need one component to fit the collapsed state but rather two components. A weak component may be related to intranetwork fields peaking around 1100 G and a stronger component probably related to network fields peaking around 1300 G. With this in mind, the explanation for the range of results may be the different contributions of the network/intranetwork components in the four data sets. 

Nevertheless, at this point in our analysis we can conclude that the MBP field strength distribution follows log-normal distributions with maxima between 1100 G and 1400 G. These values are within theoretical predictions of the convective collapse model \citep[e.g.][]{1976SoPh...50..269S}.

\begin{figure}
\begin{center}
\includegraphics[width=0.48\textwidth]{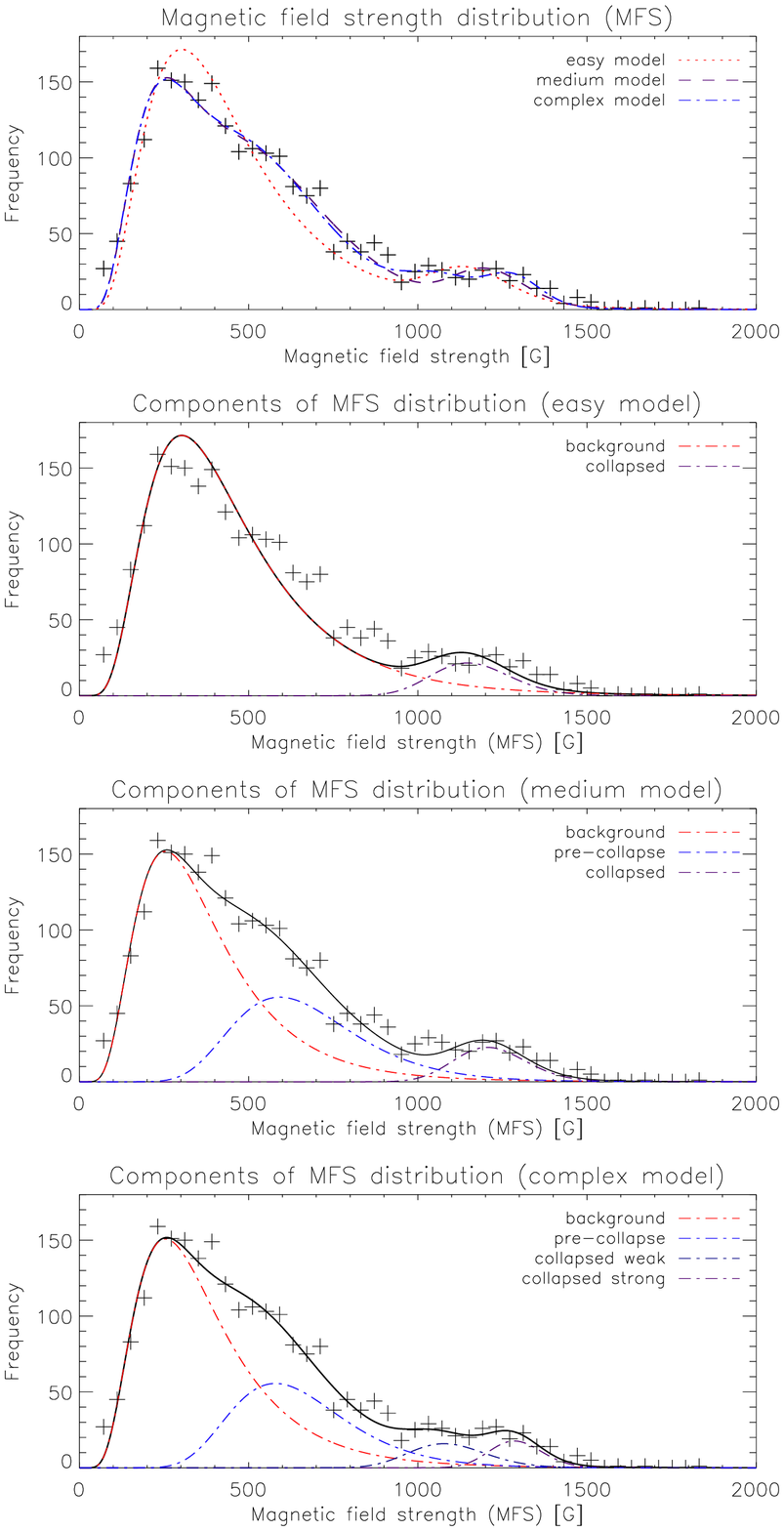}
\end{center}
\caption{Same as in Fig. \ref{figure8} but for data set IV. The goodness-of-fit value $\chi^2$ is 4.4 for the simplistic model and improves over 2.1 for the medium complex one to a final value of 1.9 for the sophisticated model.}
\label{figure10}
\end{figure}

\begin{figure}
\begin{center}
\includegraphics[width=0.48\textwidth]{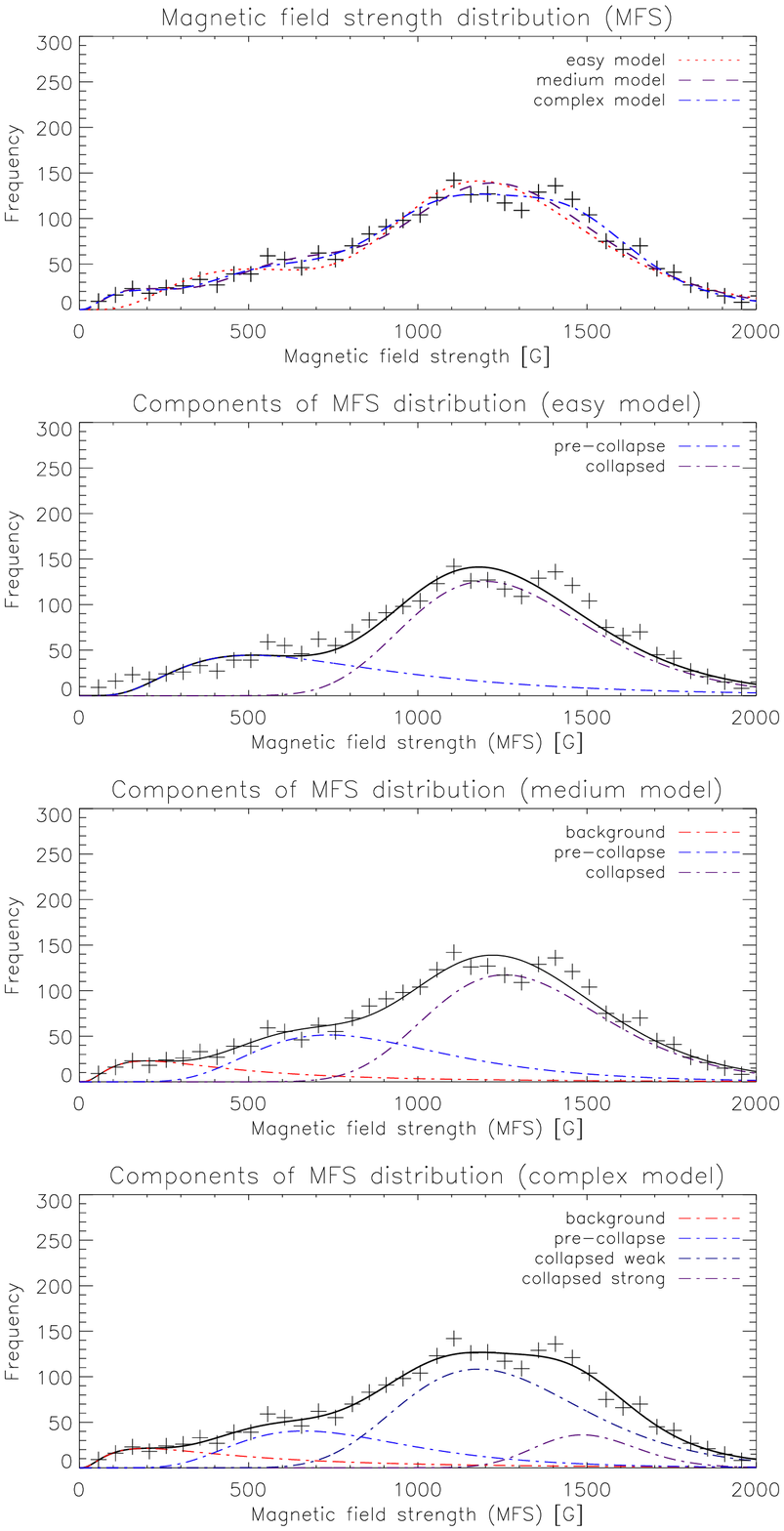}
\end{center}
\caption{Same as in Fig. \ref{figure8} but for data set II. The simplistic fit gives a rather good agreement with the measurements (reduced $\chi^2$ of 3.0); the medium complex model and the sophisticated model fit the data better with values of 1.5 and 1.1, respectively.}
\label{figure11}
\end{figure}

\section{Discussion}

We note that we are discussing magnetic field strengths in this paper, while other authors state that observations can only lead to the magnetic flux contained within a pixel. This means that the true magnetic field strength would depend on the filling factor. We implicitly assume a filling factor for MBPs of one, i.e. that the pixel is filled completely with the magnetic field. This may be justified by the fact that MBPs stretch over several pixels and by the congruence between theoretical predictions and these observations (the magnetic field strength value after the convective collapse of about 1300 G). On the other hand, it may explain how it is possible that many of the MBPs are measured with lower magnetic field strengths except of the active region data set, implying that the filling factor is thus in reality smaller than one and that the true structures are smaller and/or fragmented and yet not resolved. Nevertheless, these are speculations, so we will have to wait for future instruments with higher resolutions to bring the true physical nature to light. A possible extension of this study can hopefully be done in the future with the IMaX instrument \citep[for more details see][]{2011SoPh..268...57M} onboard the balloon-borne Sunrise mission. \citet{2010ApJ...723L.164L} has shown in a recent study that quiet Sun magnetic fields can be resolved with this instrument.

In all cases we were able to fit the background magnetic field distribution with a log-normal distribution. This distribution consists of still uncollapsed flux tubes, intermediate states in the evolution of the flux tubes, and a low fraction of wrongly identified features (identification errors). Log-normal distributions occur naturally by physical mechanisms involved in generating of the magnetic field. On the other hand, it is thinkable that the background distribution is also generated to a certain extent by measurement noise\footnote{This noise is composed of the photon noise (the higher the spectral resolution the lower the photon yield) and the noise induced by the exposure time: the longer the exposure time the better the photon statistics but the worse the contrast of the image due to the temporal averaging over the dynamics of the features contained within the FOV.}. Especially the position and height of the peak of the background distribution, which lies in our case between 100 G and 200 G, is quite often reported as a noise artefact, and it shifts with image quality \citep[see e.g.][]{2010A&A...517A..37S}. \citet{2010A&A...517A..37S} states, for example, that if one corrected the influence of noise on the measurements of quiet Sun magnetic fields, there would be no peak at all, and the probability distribution should rise very strongly for field strengths approaching zero. Furthermore, it cannot be excluded that this peak at about 190 G is caused by the interaction of the detectability (image noise) of magnetic fields and the inversion codes, which transform the polarisation signals in magnetic field vectors.

\paragraph{Data alignment:}
The alignment of filtergram data with spectro-polarimetric data inhibits an intrinsic temporal alignment problem. This is due to the design of the instrument as slit spectrograph. Therefore only data from a tiny portion of the Sun (one slit width) can be gained instantly. To obtain a suitable FOV, the slit must be moved over the surface to be mapped. This scanning process takes some time depending on the size of the FOV. This problem can be circumvented by other instrumental designs \citep[see e.g. the Sunrise mission with the IMaX instrument;][]{2011SoPh..268....1B,2011SoPh..268...57M}, which naturally inhibits other drawbacks (such as lower spectral resolution). Magnetograms taken by a slit spectrograph (like Hinode/SOT/SP) therefore show a temporal evolution of the Sun's surface.

\begin{table*}
	\centering
	\caption{Parameters for the medium complex fitting model for the magnetic field strength distribution fits as shown in Figs. \ref{figure8}, \ref{figure9}, \ref{figure10}, and \ref{figure11}. ``Data set'' specifies the used data set; $\omega$, $\sigma$, and $\mu$ specify the log-normal components; and $\chi^2$ represents the goodness-of-fit value. The indexing number indicates the corresponding component. The parameters of this table should be compared with the parameters of the simplistic model (as stated in Tables \ref{fit_1_values}, \ref{table_active}, \ref{fit_values_quiet_sun}, \ref{fit_values_quiet_sun_HR}) or the sophisticated model as summarised in Table \ref{Table7}.}
		\begin{tabular}{ccccccccccc}
		\hline
\hline
			data set  & $\omega_1$ & $\mu_1$ & $\sigma_1$   & $\omega_2$ & $\mu_2$ & $\sigma_2$  & $\omega_3$ & $\mu_3$ & $\sigma_3$ & $\chi^2$ \\
		 \hline
sunspot &  54 $\pm$ 2 & 5.46 $\pm$ 0.04 & 0.80 $\pm$ 0.03 &  25 $\pm$ 2 &  7.20 $\pm$ 0.03 &  0.13 $\pm$ 0.01 &  21 $\pm$ 2 &  6.72 $\pm$ 0.03 &  0.31 $\pm$ 0.03 & 1.3 \\
active &  10 $\pm$ 2 & 6.0 $\pm$ 0.2 & 0.8 $\pm$ 0.1 &  60 $\pm$ 3 &  7.18 $\pm$ 0.01 &  0.21 $\pm$ 0.01 &  30 $\pm$ 3 &  6.74 $\pm$ 0.04 &  0.37 $\pm$ 0.03 & 1.5\\
quiet &  51 $\pm$ 2 & 5.25 $\pm$ 0.04 & 0.67 $\pm$ 0.02 &  18 $\pm$ 1 &  7.212 $\pm$ 0.003 &  0.083 $\pm$ 0.003 & 31 $\pm$ 2 &  6.49 $\pm$ 0.02 &  0.39 $\pm$ 0.02 & 3.2 \\
quiet HR &  64 $\pm$ 3 & 5.80 $\pm$ 0.03 & 0.51 $\pm$ 0.02 &  7 $\pm$ 1 & 7.10 $\pm$ 0.01 &  0.09 $\pm$ 0.01 &  29 $\pm$ 3 &  6.47 $\pm$ 0.03 &  0.30 $\pm$ 0.02 & 2.1\\
\hline
averaged &  --  & 5.6 $\pm$ 0.3 & 0.7 $\pm$ 0.1 &  --  &  7.17 $\pm$ 0.05 &  0.12 $\pm$ 0.06 &  --  &  6.61 $\pm$ 0.15 &  0.34 $\pm$ 0.05 & -- \\
\hline
		\end{tabular}
		\label{Table6}
\end{table*}

\begin{table*}
	\centering
	\caption{Parameters for the sophisticated magnetic field strength distribution fits as shown in Fig. \ref{figure8}, \ref{figure9}, \ref{figure10}, and \ref{figure11}. ``Data set'' specifies the used data set, $\chi^2$ the goodness of fit, $\omega$, $\sigma$ and $\mu$ specifies the log-normal components. This table should be compared with the previous tables of the simplistic model (\ref{fit_1_values}, \ref{table_active}, \ref{fit_values_quiet_sun}, \ref{fit_values_quiet_sun_HR}) or the intermediate model (see Table \ref{Table6}).}
		\begin{tabular}{cccccccccc}
		\hline
\hline
			data set & $\chi^2$ & component & $\omega$ & $\mu$ & $\sigma$  & component & $\omega$ & $\mu$ & $\sigma$ \\
		 \hline
sunspot & 0.8 & 1 & 48 $\pm$ 5 & 5.39 $\pm$ 0.09 & 0.77 $\pm$ 0.04 & 2 & 20 $\pm$ 5 &  7.24 $\pm$ 0.01 &  0.11 $\pm$ 0.01 \\
 & & 3 & 18 $\pm$ 7 & 6.5 $\pm$ 0.1 & 0.4 $\pm$ 0.1 & 4 & 14 $\pm$ 7 &  6.97 $\pm$ 0.07 &  0.18 $\pm$ 0.05 \\
\hline
active & 1.1 & 1 & 11 $\pm$ 20 & 6 $\pm$ 1 & 0.9 $\pm$ 0.3 & 2 & 10 $\pm$ 6 &  7.31 $\pm$ 0.02 &  0.10 $\pm$ 0.03 \\
 & & 3 & 21 $\pm$ 9 & 6.6 $\pm$ 0.3 & 0.4 $\pm$ 0.2 & 4 & 58 $\pm$ 14 &  7.12 $\pm$ 0.03 &  0.23 $\pm$ 0.02 \\
\hline
 quiet & 1.8 & 1 & 43 $\pm$ 2 & 5.12 $\pm$ 0.04 & 0.62 $\pm$ 0.02 & 2 & 16 $\pm$ 2 &  7.23 $\pm$ 0.01 &  0.073 $\pm$ 0.004 \\
 & & 3 & 36 $\pm$ 1 & 6.33 $\pm$ 0.03 & 0.42 $\pm$ 0.03 & 4 & 8 $\pm$ 3 &  7.00 $\pm$ 0.04 &  0.14 $\pm$ 0.04 \\
\hline
 quiet HR & 1.9 & 1 & 64 $\pm$ 3 & 5.80 $\pm$ 0.03 & 0.51 $\pm$ 0.02 & 2 & 4 $\pm$ 1 &  7.16 $\pm$ 0.01 &  0.06 $\pm$ 0.01 \\
 & & 3 & 27 $\pm$ 3 & 6.44 $\pm$ 0.03 & 0.29 $\pm$ 0.03 & 4 & 5 $\pm$ 2 &  6.99 $\pm$ 0.03 &  0.10 $\pm$ 0.03 \\
\hline
averaged & -- & 1 & -- & 5.6 $\pm$ 0.4 & 0.7 $\pm$ 0.2 & 2 & -- &  7.2 $\pm$ 0.1 &  0.09 $\pm$ 0.02 \\
 values &  &3 & -- & 6.5 $\pm$ 0.1 & 0.36 $\pm$ 0.06 & 4 & -- &  7.02 $\pm$ 0.07 &  0.16 $\pm$ 0.05 \\
\hline
		\end{tabular}
		\label{Table7}
\end{table*}

For large and extended magnetic-field configurations, such as sunspots and active regions, this may not be a problem due to a slower evolution of the magnetic field \citep[see e.g.][who state that the lifetime of active regions range from days to months]{2004A&A...425..309S}, but for small scale magnetic fields \citep[see e.g. a comparison between plage and quiet Sun,][]{1992ApJ...393..782T}, which evolve in the range of minutes or even faster, this certainly is one. This problem can be dealt with in two ways. One way to solve this shortcoming is that one reduces the available FOV (as for data sets II, III \& IV) to get a better correspondence between filtergram data and magnetogram data. A smaller FOV can be scanned faster, and also the cadence of the spectro-polarimetric scans can be increased. On the other hand, a smaller FOV gives a statistically less robust result than a larger FOV would give. The other possibility would be to take several filtergram images of the whole FOV and take only a stripe of the image that is co-temporal with a certain stripe of the magnetogram. This procedure would be suitable for data set I, but would cause a major (not practicable) effort of data alignment since one should keep in mind that the stripes should be aligned on sub-pixel level, which gets more and more complicated, the smaller the FOV becomes (less alignment information\footnote{Less alignment information implies more noise and errors introduced in the analysis by increased alignment errors.}).

Therefore we decided to take one co-aligned filtergram and compare it with the full magnetogram. This gives just a rough estimate of the true magnetic field strength distribution caused by the temporal shortcomings mentioned before and by the fact that the evolution of MBPs is fast and that they show a short lifetime in the range of minutes \citep[e.g.][]{2005A&A...441.1183D,2010A&A...511A..39U}. On the other hand, the result for data set I was verified by the three other, independently taken data sets. These congruent results are gained despite the poor temporal coincidence between magnetic field information and filtergrams of data set I. This obvious antagonism leads us to the conjecture that the underlying magnetic field changes on a different (much larger) characteristic timescale than the MBP features do. 
 This can be also concluded by comparing two works of \citet{2005A&A...441.1183D,2008ApJ...684.1469D}, who investigated in one study the lifetime of MBP features deriving a mean lifetime of 3.5 min and in the other study the authors investigated patches of magnetic fields and found longer lifetimes with a mean of about 10 min. These observational constraints give rise to the idea that MBPs are formed over pre-existing magnetic field patches. These field patches are more extended than the associated MBP that increases the probability of measuring a magnetic field strength signal even after a certain time lapse. After a typical lifetime in the range of minutes, MBPs dissolve and can reappear later on at the same spot. Whether this reoccurrence really happens has to be investigated in more detail in future works. For data set I (sunspot group), we expect strong magnetic fields and large magnetic fluxes (sunspot fields), thus the underlying magnetic field may be even more stable (in the range of an hour), and several recurrent MBPs might be expected. This may explain the good agreement of the results for the different data sets.


\paragraph{Probability distributions:}
In this paper we employed a quite simple probability distribution with six free parameters consisting of two log-normal distributions in a first step. As a result the derived parameters are two geometrical mean values $\mu$, two multiplicative standard deviations $\sigma$, and two weighting coefficients $\omega$. As one can see by the corresponding distributions (Figs. \ref{figure2}, \ref{figure4}, \ref{figure6}, and \ref{figure7}), this simple model does not fit the observed distribution perfectly, which can also be concluded from the corresponding reduced $\chi^2$ values, which show values between 3.0 and 8.3. The partly high $\chi^2$ values can be explained e.g. by the impossibility of such an easy fitting model to differentiate and fit the two distinct bumps at about 1000 G and 1500 G, respectively. In the following we discuss more complicated fitting models by adding first another (third) log-normal component (to enhance the fitting for intermediate fields) and finally a fourth log-normal component (splitting the collapsed state into a weak and a strong field case). Furthermore we want to offer some possible explanation for the appearance of these components.

Starting with data set I and applying the medium-complex fitting model on the results (three log-normal components) shown in the third panel from top of Fig. \ref{figure8}, we see that the agreement between the measured distribution and the fitted curve is already better when compared to the simplistic model (given in the second panel from top). The solid black line shows the resulting total fit, while the single components are shown in color (red for the backrgound, blue for the pre-collapse or intermediate state, and purple for the collapsed MBP state). The quality of the fit improved from a value of 3.9 to 1.3. The figure also shows that the fitted distribution still misses a major feature around 1000 G. To accomplish a well-fitted model distribution, we finally added a fourth component. The plot of this four-component model is shown in the bottom panel. One can see that the four-component model fits the measured distribution nearly perfectly. This is also expressed by the fit-quality measure, namely the reduced $\chi^2$ that now has a value of 0.8.

Now the question arises whether we can assign an intuitive meaning to the three- and four-component models, respectively.
We tried to do so and came up with the following interpretations for the components (see also the legend in the figure):
\begin{itemize}
	\item background: represents probably uncollapsed fields and purely background magnetic fields related to the acting solar dynamo processes; shows a geometric mean value of ($270~\cdot|\div~2 $) G.
	\item pre-collapse: consists of MBPs in their pre- and post-collapse phase (an intermediate stage in which background magnetic fields get unstable and finally reach the collapsed state and a new equilibrium; obtained geometric mean value of ($670~\cdot|\div~1.4 $) G).
	\item weak collapsed fields: the collapsed state of flux tubes. We assume that the weak component refers to collapsed intranetwork fields (estimated geometric mean value of ($1120~\cdot|\div~1.2 $) G).
	\item strong collapsed fields: in the case of the four-component model, the collapsed state seems to split up in a weak and a strong component. The strong component probably refers to collapsed network magnetic fields featuring a geometric mean value of ($1340~\cdot|\div~1.1 $) G.
\end{itemize}
From this interpretation it is clear that the contribution of the components in each single data set may vary depending on the kind of observed solar surface (e.g. exhibiting more or less magnetic network). Nevertheless, if the suggested four-component model is indeed related to true physical processes, it should be possible to fit all four data sets with the same model. Furthermore, the fitted distribution functions should yield nearly the same parameters, and only the weighting of the components should change from case to case. To test this we applied the medium complex and sophisticated model also to all of the other data sets (shown in Figs. \ref{figure9}, \ref{figure10}, and \ref{figure11}).

It is indeed possible to fit all four data sets very well with the three- and four-component models. The quality of the fit increases (see the figure captions), as can be supposed due to the higher number of fitting parameters. Only in the case of data set II also the simplistic fit gives a good agreement with the measurements expressed by reduced $\chi^2$ of 3.0. Nevertheless also in this case the medium complex and the sophisticated model fit better by far. An overview of these parameters can be found in Tables \ref{Table6} and \ref{Table7}. One can see that the parameters describing the different components are obtained in all cases with very similiar values. This is an indication that the distribution actually consists of several components and that the easier model misses the weaker components\footnote{If components would not be required or randomly fitted into the distributions, they should not be assigned similiar values for each data set but would instead turn out to be assigned different values for each case.}.

We can conclude that the two component model already reasonably fits the observed data set, and yet a three- or four-component model fits the observed distributions significantly better. The necessity of the additional components can be indirectly verified by the small error in the weighting parameters (see Tables \ref{Table6} and \ref{Table7}). In case of overfitting or introducing  unnecessary components to a fit, the error in the weighting parameter should be in the range of the weight itself. Furthermore, we can assign by intuition a distinct meaning to each component of the complex models. Nevertheless, whether such a model makes sense would have to be proven by theoretical means and by investigations of single MBPs in more detail; i.e., do they form out of a background magnetic field and go through an intermediate state to finally reach the collapsed state? What happens thereafter, and can they reach two different distinguishable
collapsed states (associated to network and intranetwork fields)? 

In the future more detailed investigations will either prove the physical significance of these complex models or show that they are rather complicated functions that fit the observed distributions without any physical meaning buried in their components. 

Finally, we stress that our sophisticated fit model is highly non-linear, and thus small changes in the starting conditions can change the outcome for the components (especially the weak ones). It is therefore necessary to give an educated guess about the right starting parameters. This means, e.g., visually identifying the peaks in the distribution and setting the starting parameters for the geometrical mean value of the distributions accordingly, i.e the value should be centred on the peak. Nevertheless, the main result (the bump around 1300 G in the magnetic field distribution for MBPs) can be deduced from the sophisticated models, as well as from the easier model, and has to be taken as evident.

\section{Conclusions}

We analysed four different data sets of Hinode/SOT comprising G-band and SP data. The four data sets show different solar surface features (active region with sunspots and pores, and quiet Sun). The magnetic field strength distribution of MBPs was obtained for those four data sets by comparing fully inverted SP data (magnetograms) with carefully co-aligned G-band filtergrams.
All the data sets showed the same magnetic field strength distribution behavior with a strong peak around 1300 G. This value was theoretically predicted by the convective collapse model \citep[as e.g. stated by ][]{1976SoPh...50..269S} and is now observationally verified in this paper. Furthermore, we showed that the magnetic field background signal (non-collapsed flux tubes), as well as the complete distribution, can be fitted by log-normal distribution components. The number of components for the model and the interpretation are outstanding, but probably linked to the dynamo and formation processes that form the magnetic fields/MBPs. The log normal component related to the collapsed flux tubes shows a contribution ranging from 10\% to 30\% (except for the active region case where it reaches 60\%), indicating that a large fraction of MBPs exhibit weak and uncollapsed magnetic fields.

\begin{acknowledgements}
The research was funded by the Austrian Science Fund (FWF): P20762, P23618, and J3176. We are
grateful to the Hinode team for the
possibility of using their data.
Hinode is a Japanese mission developed and launched by ISAS/JAXA, collaborating with NAOJ as a domestic partner, and with NASA and STFC (UK) as international partners. Scientific operation of the Hinode mission is conducted by the Hinode science team organised at ISAS/JAXA. This team mainly consists of scientists from institutes in the partner countries. Support for the post-launch operation is provided by JAXA and NAOJ (Japan), STFC (U.K.), NASA (U.S.A.), ESA, and NSC (Norway).
Hinode SOT/SP Inversions were conducted at NCAR in the framework of the Community Spectro-polarimtetric Analysis Center (CSAC; http://www.csac.hao.ucar.edu/). D. U., A. H. and O. K. are grateful to
the {\"O}AD ({\"O}sterreichischer Austauschdienst) for financing a
scientific stay at the Pic du Midi Observatory. M.R. is grateful to the Minist\`{e}re des Affaires Etrang\`{e}res et Europ\'{e}ennes for financing a stay at the University of Graz. In addition D.U. wants to thank the {\"O}AD and M\v{S}MT again for financing a
short research stay at the Astronomical Institute of the Czech Academy of
Sciences in Ondrejov in the framework of the project MEB061109. Furthermore, J.J. wishes to express his gratitude to the {\"O}AD and M\v{S}MT for financing a stay at the University of Graz. Finally, J.J. gratefully acknowledges the support from GA~CR~P209/12/0287 and RVO:67985815. Last but not least, we also thank the anonymous referee whose comments and remarks helped us to improve the paper.

\end{acknowledgements}
\bibliographystyle{aa}
\bibliography{Literaturverzeichnis}

\end{document}